\documentclass[aps,twocolumn,showpacs,preprintnumbers,nofootinbib,prd,superscriptaddress,groupedaddress,10pt]{revtex4-1}

\makeatletter
\def\l@subsubsection#1#2{}
\def\l@subsubsubsection#1#2{}
\makeatother

\usepackage{graphicx,amssymb,amsmath,amsthm,amsfonts,epsfig,epsf}
\usepackage[usenames]{color}
\usepackage{epstopdf}

\usepackage{aas_macros}
\usepackage{bm}
\usepackage{dcolumn}
\usepackage{latexsym}
\usepackage{rotating}
\usepackage{longtable}

\usepackage{enumerate}
\usepackage{tensor,multirow}
\usepackage{url}
\usepackage[linktocpage]{hyperref}

\def\nn{\nonumber}

\begin{document}

\title{On gravitational-wave echoes from neutron-star binary coalescences}
%%%%

\author{Paolo Pani}\email{paolo.pani@roma1.infn.it}
\affiliation{Dipartimento di Fisica, ``Sapienza'' Universit\`a di Roma \& Sezione INFN Roma1, Piazzale Aldo Moro 5, 00185, Roma, Italy}
\author{Valeria Ferrari}\email{valeria.ferrari@roma1.infn.it}
\affiliation{Dipartimento di Fisica, ``Sapienza'' Universit\`a di Roma \& Sezione INFN Roma1, Piazzale Aldo Moro 5, 00185, Roma, Italy}

\begin{abstract} 
A tentative detection of gravitational-wave echoes in the post-merger signal of GW170817 has been recently claimed  at $4.2\sigma$ significance level. It has been speculated that the signal might provide evidence for near-horizon quantum structures in the remnant exotic object. We point out that if the remnant object is an ultracompact neutron star, echoes are expected for objects with radius only slightly smaller than that of an ordinary neutron star. The reported echoes at $\approx72\,{\rm Hz}$ are compatible with a toy model of incompressible star with mass approximately $M\in(2,3) M_\odot$ and radius close to the Buchdahl limit, $R\approx 9GM/(4c^2)$. If confirmed, low-frequency gravitational-wave echoes would be in tension with all current neutron-star models and would have dramatic implications for nuclear physics and gravity.
\end{abstract}

\maketitle

%%%%%%%%%%%%%%%%%%%%%%%%%%%%%%%%%%%%%%%%%%%%%%%%%%%%%%%%%%%%%%%%%%%%%%%%%%%%%
\noindent{\bf{\em Introduction.}}
%%%%%%%%%%%%%%%%%%%%%%%%%%%%%%%%%%%%%%%%%%%%%%%%%%%%%%%%%%%%%%%%%%%%%%%%%%%%%
It has been recently suggested that gravitational-wave~(GW) echoes~\cite{Cardoso:2016rao,Cardoso:2016oxy} in the post-merger GW signal from a binary coalescence might be a generic feature of quantum corrections at the horizon scale, and might provide a smoking-gun signature of exotic compact objects (see~\cite{Cardoso:2017cqb,Cardoso:2017njb} for a review).
In the last two years, tentative evidence for echoes in the combined LIGO/Virgo binary black-hole~(BH) events have been reported~\cite{Abedi:2016hgu,Conklin:2017lwb} with controversial results~\cite{Ashton:2016xff,Abedi:2017isz,Westerweck:2017hus,Abedi:2018pst}. This has also motivated several studies on the modeling of the echo waveform~\cite{Nakano:2017fvh,Mark:2017dnq,Bueno:2017hyj,Maselli:2017tfq,Wang:2018mlp,Correia:2018apm,Wang:2018gin}. Very recently, a tentative detection of echoes in the post-merger signal of neutron-star (NS) binary coalescence GW170817~\cite{TheLIGOScientific:2017qsa} has been claimed at $4.2\sigma$ significance level~\cite{Abedi:2018npz}. It has been speculated that these GW echoes could be related to the quantum properties of an exotic ``BH remnant''.

If GW echoes are confirmed in the event GW170817, it is of utmost importance to understand whether they are related to new physics at the horizon scale~\cite{Cardoso:2017cqb,Cardoso:2017njb} or if similar repeated signals might arise also in other scenarios, for example as low-frequency quasiperiodic oscillations in the post-merger environment~\cite{Abedi:2018npz}, or as quasinormal modes~\cite{Kokkotas:1999bd} of a remnant (possibly exotic) star.

\begin{figure}[th]
\centering
\includegraphics[width=0.43\textwidth]{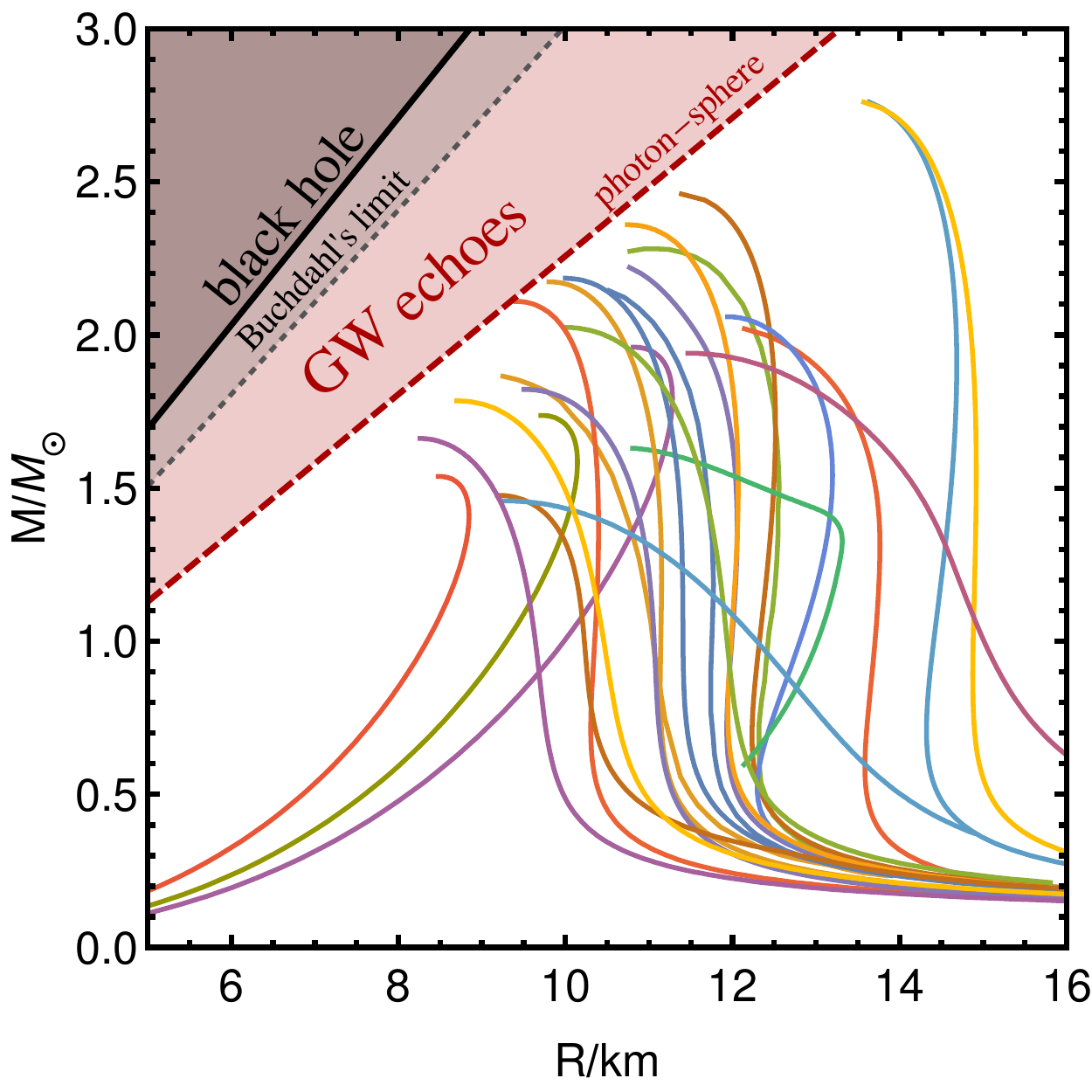}
\caption{Mass-radius diagram of nonspinning NSs for several representative equations of state~\cite{Lattimer:2006xb,Ozel:2016oaf} (data taken from~\cite{OzelFreireWeb}). GW echoes require the star to feature a photon-sphere (red shaded area) and low-frequency echoes require configurations located far deep into this region. Echoes in the post-merger signal would imply that the remnant is more compact than a NS with an ordinary equation of state.
}
\label{fig:MR}
\end{figure}

In this letter we point out that GW echoes are not a prerogative of quantum corrections at the horizon scale. Similar signals have been long known to arise in ultracompact stars~\cite{Ferrari:2000sr,Kokkotas:2000is} featuring a photon-sphere~\cite{Cardoso:2014sna,Cardoso:2017cqb,Cardoso:2017njb}, the latter being able to effectively trap radiation within the stellar interior~\cite{1991RSPSA.434..449C,Ferrari:2000sr} (for a mathematical discussion on photon surfaces in general relativity, see~\cite{Claudel:2000yi}).
We show that echoes at the reported frequency $f_{\rm echo}\approx72\,{\rm Hz}$~\cite{Abedi:2018npz} would arise naturally if the remnant object is an ultracompact star only slightly more compact than an ordinary NS, and are thus not necessarily related to Planckian corrections at the horizon scale. In the static case, the existence of a photon-sphere requires $R<3M$, where $M$ and $R$ are the stellar mass and radius, respectively (henceforth, we use $G=c=1$ units). Ordinary equations of state do not support self-gravitating configurations which are so compact~\cite{Lattimer:2006xb,Ozel:2016oaf} (see Fig.~\ref{fig:MR}). Therefore, should echoes be confirmed in GW170817, they might provide evidence for a very exotic state of matter formed after the merger, but they are not necessarily associated with near-horizon quantum structures.

%

%%%%%%%%%%%%%%%%%%%%%%%%%%%%%%%%%%%%%%%%%%%%%%%%%%%%%%%%%%%%%%%%%%%%%%%%%%%%%
\noindent{\bf{\em Echoes from ultracompact stars.}}
%%%%%%%%%%%%%%%%%%%%%%%%%%%%%%%%%%%%%%%%%%%%%%%%%%%%%%%%%%%%%%%%%%%%%%%%%%%%%
As an illustrative model, we consider an incompressible, constant-density NS in general relativity, described by the Schwarzschild solution~\cite{1999physics..12033S,Wald}. In the static case, the metric reads $ds^2=-e^{2\Phi}dt^2+\frac{dr^2}{1-2{\cal M}(r)/r}+r^2d\Omega^2$, with
\begin{eqnarray}
{\cal M}&=&\frac{4\pi}{3}\rho r^3\,,\,\,\,P=\rho\left(\frac{\sqrt{1-2Mr^2/R^3}-\sqrt{1-2M/R}}{3\sqrt{1-2M/R}-\sqrt{1-2Mr^2/R^3}}\right)\,,\nonumber\\
e^{\Phi}&=&\frac{3}{2}\sqrt{1-2M/R}-\frac{1}{2}\sqrt{1-2Mr^2/R^3}\,,
\end{eqnarray}
when $r<R$. Here $P$ and $\rho=3M/(4\pi R^3)$ are the pressure and the density of the fluid. When $r>R$ the metric reduces to the Schwarzschild vacuum solution, ${\cal M}(r)=M$, $e^{2\Phi}=1-2M/r$, $P=\rho=0$.
The minimum radius of the object is $R_B\equiv 9/4M$~\cite{Buchdahl:1959zz}, so that the external spacetime can feature a photon-sphere when $R_B<R<3M$. 

The characteristic echo time scale is related to the light crossing time from the center of the star to the photon-sphere,
%%%
\begin{equation}
 \tau_{\rm echo}=\int_{0}^{3M} dr\frac{1}{\sqrt{e^{2\Phi}(1-2{\cal M}/r)}}\,, \label{tauecho}
\end{equation}
%%%%
and the echo frequency is well approximated by the roundtrip frequency, $f_{\rm echo}\sim \pi/\tau_{\rm echo}$~\cite{Cardoso:2016rao,Cardoso:2016oxy,Cardoso:2017cqb,Cardoso:2017njb}.

Models of quantum corrections at the horizon scales typically assume that $R\sim2M+l_p$, where $l_p\ll M$ is the Planck length. In this case, the echo delay time reduces to the BH ``scrambling''~\cite{Hayden:2007cs,Sekino:2008he} time, $\tau_{\rm echo}\sim M|\log(l_p/M)|$, irrespectively of the object interior~\cite{Cardoso:2016oxy}. This is the case studied so far for wormholes~\cite{Cardoso:2016rao}, gravastars~\cite{Cardoso:2016oxy}, Kerr-like objects with a reflective surface~\cite{Cardoso:2016oxy,Abedi:2016hgu,Maggio:2017ivp}, and other quantum-dressed BH-like objects~\cite{Conklin:2017lwb} (for a review, see Refs.~\cite{Cardoso:2017cqb,Cardoso:2017njb}).

For an ultracompact star the situation is dramatically different, because redshift at the surface is moderate and most of the crossing time accumulates in the \emph{interior} of the object. By defining $\epsilon=R/R_B-1$, it is straightforward to obtain
%%%
\begin{eqnarray}
 \frac{\tau_{\rm echo}}{M} &=& \frac{27 (\epsilon +1)^2}{16 \sqrt{\epsilon }} \left(\cot ^{-1}\left(\sqrt{\epsilon }\right)+\cot ^{-1}\left(3 \sqrt{\epsilon }\right)\right)\nn\\
 &+&\frac{9 \epsilon -3}{4} +2 \log \left(\frac{9 \epsilon +1}{4}\right)\sim \frac{27\pi}{16}\epsilon^{-1/2}\,,
\end{eqnarray}
where the first and second lines come from the internal ($0<r<R$) and external $(R<r<3M$) contributions to the integral~\eqref{tauecho}, respectively. The last approximation is valid when $\epsilon\to0$ and depends only on the internal contribution. The echo delay time diverges in the Buchdahl's limit, but not logarithmically as for near-horizon quantum corrections.

Therefore, for a nonspinning constant-density star, the echo frequency reads
%%%
\begin{equation}
 f_{\rm echo}\sim \frac{\pi}{\tau_{\rm echo}}\approx 46\left(\frac{2.7 M_\odot}{M}\right)\left(\frac{\epsilon}{10^{-6}}\right)^{1/2}\,{\rm Hz}\,.\label{frequency}
\end{equation}
%%%
For $\epsilon\sim {\cal O}(10^{-6}-10^{-5})$, the above frequency is comparable to what reported in Ref.~\cite{Abedi:2018npz} for GW170817.
In this case, $R-2M\approx \left(\frac{M}{2.7M_\odot}\right){\rm km}$, i.e. the difference between the stellar radius and the Schwarzschild radius is macroscopic, not Planckian. 
When $\epsilon\ll1$, Eq.~\eqref{frequency} agrees with the time-domain results of Ref.~\cite{Ferrari:2000sr} and with the fundamental quasinormal mode frequency of this object.

%%%%%%%%%%%%%%%%%%%%%%%%%%%%%%%%%%%%%%%%%%%%%%%%%%%%%%%%%%%%%%%%%%%%%%%%%%%%%
\noindent{\bf{\em Discussion.}}
%%%%%%%%%%%%%%%%%%%%%%%%%%%%%%%%%%%%%%%%%%%%%%%%%%%%%%%%%%%%%%%%%%%%%%%%%%%%%
The above estimates are valid for nonrotating, constant-density stars. 
While the angular velocity $\Omega$ of ordinary NSs is expected to be small compared to the mass-shedding limit, $\Omega_K=\sqrt{M/R^3}\approx 22\left(\frac{2.7M_\odot}{M}\right)\,{\rm kHz}$ (for a star with $R\approx 9/4 M$), one cannot exclude the possibility of highly-spinning merger remnants, or the fact that exotic stars can spin faster than ordinary NSs. It is therefore relevant to include spin effects in our model.
This can be done perturbatively (i.e., for $\Omega\ll\Omega_K$) through the Hartle-Thorne formalism~\cite{Hartle:1967he,Hartle:1968si}. As in the Kerr case~\cite{Abedi:2016hgu,Wang:2018gin}, one could estimate the echo frequency through the crossing time of \emph{principal null geodesics}, i.e. those geodesics that more directly fall into the object, having angular-momentum-to-energy ratio equal to the angular velocity of the central object~\cite{Chandra}.

\begin{figure}[th]
\centering
\includegraphics[width=0.43\textwidth]{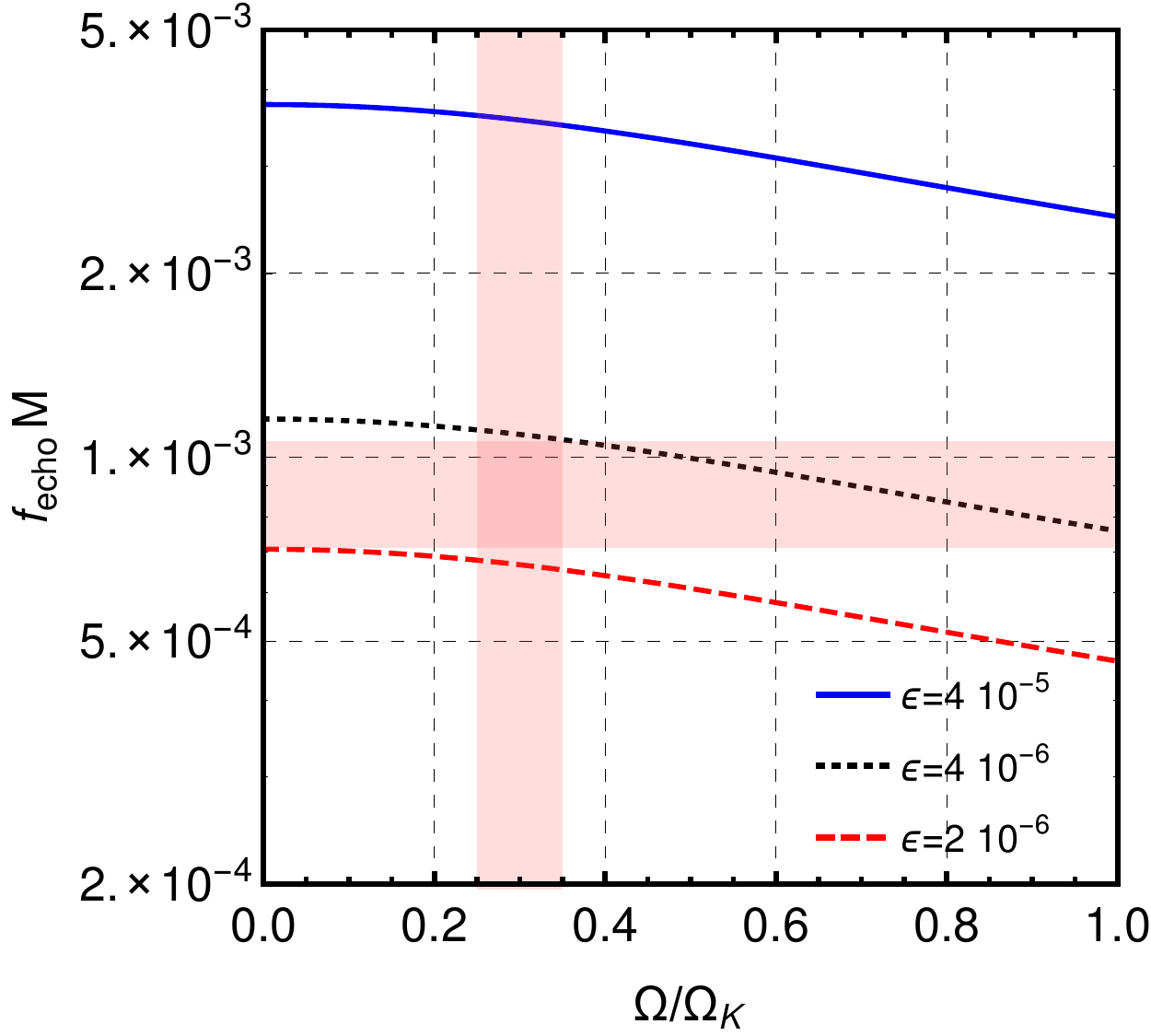}
\caption{Dimensionless echo frequency $f_{\rm echo}M$ as a function of the angular velocity $\Omega$ of the object (normalized by the mass-shedding limit, $\Omega_K=\sqrt{M/R^3}$) for different values of $\epsilon=R/R_B-1$ and to quadratic order in $\Omega/\Omega_K$. The echo frequency monotonically decreases with the spin. Shaded areas refer to a reference measurement of $M=2.5^{+0.5}_{-0.5}M_\odot$, $\Omega=0.3^{+0.05}_{-0.05}\Omega_K$, and $f_{\rm echo}\approx72\,{\rm Hz}$.
}
\label{fig:spin}
\end{figure}

Figure~\ref{fig:spin} shows the echo frequency as a function of the angular velocity $\Omega$ of the star for several values of $\epsilon$, as computed by solving the Tolman-Oppenheimer-Volkoff equations numerically to second order in the spin and computing the principal null geodesics on the equator of this metric. The echo frequency decreases monotonically with the spin, because spinning configurations can be more compact than their nonspinning counterpart. Thus, although low-frequency echoes can be explained also by nonspinning models [see Eq.~\eqref{frequency}], including the angular momentum would permit to match the same echo frequency with a less compact configuration.

Ultracompact exotic objects are unstable against the ergoregion instability~\cite{1978CMaPh..63..243F,Moschidis:2016zjy,Brito:2015oca} when they spin sufficiently fast~\cite{1978RSPSA.364..211C,1996MNRAS.282..580Y,Cardoso:2007az,Cardoso:2008kj,Chirenti:2008pf}.
However, the instability time scale $\tau_{\rm ergoregion}\gg M$ (see, e.g., Fig.~4.19 in Ref.~\cite{Brito:2015oca}) and is comparable to (and typically longer than) $\tau_{\rm echo}$~\cite{Maggio:2017ivp}. Thus, unless the instability is quenched by some other mechanism~\cite{Maggio:2017ivp}, an ultracompact spinning star formed after the merger should produce echoes while losing angular momentum over a time scale much longer than the prompt ringdown time scale. In this scenario, the final stationary configuration is expected to be a slowly-spinning (or less compact) exotic star or a BH~\cite{Cardoso:2014sna}.

Incompressible, constant-density stars are just a toy model, since they predict infinite speed of sound in the fluid. However, it is reasonable to expect that similar results would apply to more realistic models of ultracompact stars featuring a photon-sphere and near their corresponding maximum compactness. In particular, GW echoes would imply an equation of state that is close enough to that of an incompressible fluid.
Remarkably, none of the ordinary equations of state describing the NS core can support such ultracompact self-gravitating configurations~\cite{Lattimer:2006xb,Ozel:2016oaf} (see Fig.~\ref{fig:MR}). Strange stars with large values of the bag constant marginally feature a photon-sphere, although the maximum compactness of stable configurations is not enough to explain echoes at ${\cal O}(100){\rm Hz}$~\cite{Mannarelli:2018pjb}. In particular, low-frequency echoes are expected only for configurations laying deep within the red shaded region of Fig.~\ref{fig:MR}. No known equation of state supports compact stars in this region.

The ``echoing remnant'' does not need to be stable, at least as long as its instability time scale is sufficiently long. Thus, it would be also interesting to investigate the presence of a photon-sphere in configurations that are usually dismissed due to their instability, such as highly-spinning compact stars (possibly) in the unstable branch. 

Finally, it is interesting to compare our model of incompressible star with the current status of BH-like exotic objects~\cite{Cardoso:2017cqb,Cardoso:2017njb}. The latter are all motivated by quantum corrections at the horizon scale, so they would provide low-frequency GW echoes only when $R-2M\approx l_p$. In addition, to the best of our knowledge there are very few first-principle models of ultracompact objects. Many of these models require either thin shells of matter and junction conditions between the interior and the exterior, or violations of the energy conditions, or unrealistic assumptions such as perfectly-reflective surfaces. It is therefore unclear how could these models form dynamically. An exception in this sense are boson stars, which are described by a solid theoretical framework and can form in the gravitational collapse and in mergers~\cite{Liebling:2012fv}; however, known boson-star models are not compact enough to support echoes~\cite{Cardoso:2016oxy,Cardoso:2017cqb,Cardoso:2017njb,Palenzuela:2017kcg}.
Given the current knowledge of exotic compact objects that can support low-frequency GW echoes, a nearly-incompressible fluid star is certainly less speculative than the previous models. In fact, at the moment the model considered in this work is the only one that arises as a solution of Einstein equation in the presence of a single fluid and with no ad-hoc assumptions.

To summarize --~albeit not necessarily associated with near-horizon quantum structures as conjectured in Ref.~\cite{Abedi:2018npz}~-- GW echoes in GW170817 might be a smoking gun of the formation of a very exotic state of matter in the extremely compact remnant star, and would have dramatic implications for nuclear physics and gravity.

%%%%%%%%%%%%%%%%%%%%%%%%%%%%%%%%%%%%%%%%%%%%%%%%%%%%%%%%%%%%%%%%%%%%%%%%%%%%%
\noindent{\bf{\em Acknowledgments.}}
%%%%%%%%%%%%%%%%%%%%%%%%%%%%%%%%%%%%%%%%%%%%%%%%%%%%%%%%%%%%%%%%%%%%%%%%%%%%%
We thank Massimo Mannarelli and Thomas Sotiriou for interesting correspondence. 
P.P. acknowledges financial support provided under the European Union's H2020 ERC, Starting Grant agreement no.~DarkGRA--757480.
The authors would like to acknowledge networking support by the COST Action CA16104.
%%%%%%%%%%%%%%%%%%%%%%%%%%%%%%%%%%%%%%
% \bibliographystyle{apsrev4}
\bibliographystyle{utphys}
\bibliography{Ref}
\end{document}